\documentclass[twocolumn,showpacs,amsmath,amssymb,pre,aps,shttps://www.overleaf.com/project/61f1257d9e8b370c12f0212auperscriptaddress]{revtex4-1}

\usepackage{graphicx}
\usepackage{dcolumn}
\usepackage{bm}
\usepackage{lineno}
\usepackage{dcolumn}
\usepackage{bm}
\usepackage{hyperref}
\usepackage{mathtools}
\usepackage{appendix}
\usepackage{footnote}
\usepackage{color}
\graphicspath{ {images/} }
\usepackage{braket}
\usepackage{xcolor}

\usepackage{braket}
\newcommand{\ee}{\end{equation}}
\newcommand{\be}{\begin{equation}}
\newcommand{\ea}{\end{eqnarray}}
\newcommand{\ba}{\begin{eqnarray}}
\newcommand{\ean}{\end{eqnarray*}}
\newcommand{\ban}{\begin{eqnarray*}}


\begin{document}

\title[Hydrodynamic Bell test]{A platform for investigating Bell correlations in pilot-wave hydrodynamics}
\thanks{We acknowledge the generous financial support of the NSF through grants CMMI-1727565 and CMMI-2154151, the European Union’s Horizon 2020
Research and Innovation Program under the Marie Sklodowska-Curie project EnHydro, grant No. 841417, and the CNPq under (PQ1D) 307078/2021-3  and
FAPERJ  CNE, Project No. E-26/201.156/2021.}

\author{Konstantinos Papatryfonos}
\affiliation{Gulliver UMR CNRS 7083, ESPCI Paris, Universit\'e PSL, 75005 Paris, France} 
\affiliation{Department of Mathematics, MIT, 77 Massachusetts Ave., Cambridge MA 02139, USA}
\email{kpapatry@mit.edu}

\author{Louis Vervoort}
\affiliation{Higher School of Economics, 101000 Moscow, Russian Federation} 
\email{lvervoort@hse.ru}

\author{Andr\'e Nachbin}
\email{nachbin@impa.br}
\affiliation{Instituto de Matem\'{a}tica Pura e Aplicada, Estrada Dona Castorina 110, Rio de Janeiro, RJ, 22460-320, Brazil}

\author{Matthieu Labousse}
 \email{matthieu.labousse@espci.psl.eu}
\affiliation{Gulliver UMR CNRS 7083, ESPCI Paris, Universit\'e PSL, 75005 Paris, France} 

\author{John W.M. Bush}
 \email{bush@math.mit.edu}
\affiliation{Department of Mathematics, MIT, 77 Massachusetts Ave., Cambridge MA 02139, USA}

\date{\today}
\begin{abstract}
Since its discovery in 2005, the hydrodynamic pilot-wave system has provided a concrete macroscopic realization 
of wave-particle duality and concomitant classical analogs of a growing list of quantum effects. The question naturally arises as to whether this system might support statistical states that violate Bell’s inequality, and so yield a classical analog of quantum entanglement. We here introduce a new platform for addressing this question, a numerical model of coupled bipartite tunneling in the hydrodynamic pilot-wave system. We demonstrate that, under certain conditions, the Bell inequality is violated in a static Bell test owing to correlations induced by the wave-mediated coupling between the two subsystems. The establishment of non-factorizable 
states with two spatially separated classical particles introduces the possibility of novel forms of quantum-inspired classical computing.
\end{abstract}

\keywords{pilot-wave hydrodynamics, tunneling, wave coupling, Bell tests}
\maketitle

\noindent{\bf Significance} Millimeter-scale droplets self-propelling along the surface of a vibrating liquid bath
represent a macroscopic realization of wave-particle duality, an oddity once thought to be
exclusive to the microscopic realm. This classical pilot-wave system bears a strong 
resemblance to an early model of quantum dynamics proposed by Louis de Broglie, 
and has provided the basis for a surprising number of hydrodynamic quantum analogs.
We here take the first step towards assessing the plausibility of achieving entanglement 
with this hydrodynamic system. Our numerical investigation of the dynamics of 
two distant droplets reveals violations of Bell's inequality that may be rationalized 
in terms of the wave-induced coupling between them. The resulting non-factorizable, 
spatially separated classical states may find application in quantum-inspired classical 
computing.

\vspace*{0.1in}

In 2005, Yves Couder and Emmanuel Fort~\cite{Couder2005a,Couder2006} discovered that a millimetric droplet may self-propel along the surface of a vibrating fluid bath through a resonant interaction with its own wave field. The resulting `walker' consists of a droplet dressed in a quasi-monochromatic wave field, and represents a concrete, macroscopic example of wave-particle duality~\cite{Fort2010}. Remarkably, this hydrodynamic pilot-wave system exhibits many features previously thought to be exclusive to the microscopic, quantum realm~\cite{Bush2015a, BushOza}. Notable examples include single-particle diffraction and interference~\cite{Couder2006,Pucci2018,Ellegaard2020}, quantized orbits~\cite{Fort2010,Perrard2014}, unpredictable tunneling~\cite{Eddi2009b}, Friedel oscillations~\cite{Saenz2019a}, spin lattices~\cite{saenz_emergent_2021}, and quantum-like statistics and statistical projection effects in corrals~\cite{Harris2013a,Saenz2018b}. In all instances, the emergent quantum behavior may be rationalized in terms of the droplet’s non-Markovian pilot-wave dynamics~\cite{BushOza}. Specifically, the instantaneous wave force imparted to the drop during impact depends on the droplet's history. Thus, the drop navigates a potential landscape of its own making~\cite{BushOza}, and the hydrodynamic pilot-wave system is said to be endowed with `memory'~\cite{Eddi2011a}. 

In several settings, long-range interactions in the walking-droplet system emerge dynamically through the influence of the pilot-wave field
~\cite{BushOza}. For example, long-range lift forces are generated when a walking droplet interacts with a submerged pillar~\cite{Harris2018} or well~\cite{Saenz2019a}, and
long-range correlations between distant walkers may be established through the influence of the intervening wave field~\cite{Nachbin2018,Nachbin2022}. Recently, Papatryfonos {\it et al.}~\cite{papatryfonos} established a walking-droplet analog of superradiance, an effect originally attributed to quantum interference of two or more entangled atoms~\cite{DeVoe,makarov_metastable_2004, Kaminer}, but subsequently rationalised in terms of classical electromagnetic wave interference~\cite{TANJISUZUKI2011201}.
The totality of the quantum-like features evident in the 
hydrodynamic pilot-wave system 
naturally raises the following question. Might this walking-droplet system
provide a platform for demonstrating a classical analog of 
entanglement~\cite{de2015emerging,Vervoort2018}; specifically, might it permit violations of Bell's inequality? 

Bell's Theorem was derived by John Bell in 1964~\cite{Bell1964} with a view to informing the Bohr-Einstein debate concerning the completeness of quantum theory~\cite{Einstein1935}. Hidden variables are those variables that would be required for a complete description of quantum dynamics, including the position and momentum of a microscopic particle.
Bell tests inform what class of hidden variable theories are viable candidates for a causally complete quantum theory.
A Bell test can be performed on any probabilistic system consisting of two subsystems (A and B) on which one measures a dichotomic property $X$ (with stochastic outcomes of +1 or -1) that depends on some ‘analyzer setting’ ($\alpha$ or $\beta$). The measurement $X_A$ made at the left measurement system depends on the analyzer setting $\alpha$ which may take values $a$ or $a^{\prime}$; likewise, the measurement $(X_B)$ made at the right measurement system depends on $\beta$ which may take values $b$ or $b^{\prime}$. In the derivation of the Bell inequality (Eq. 1), it is assumed that the two subsystems undergo only local interactions;  specifically, $X_A$ depends on $\alpha$ and not $\beta$; likewise, $X_B$ depends on $\beta$ and not $\alpha$. This assumption is referred to as `Bell locality'. 
Another assumption made is that the hidden variables 
that prescribe $X$ are independent of $\alpha$ and $\beta$, a condition referred to as ‘measurement independence’.
Bell’s theorem~\cite{Bell1964} implies 
that for any classical system for which Bell locality and measurement independence hold, 
the quantity $S\left(\alpha=a,\beta=b,\alpha=a' ,\beta=b'\right)=  M(a,b) + M(a^{\prime},b) + M(a,b^{\prime})
         -M(a^{\prime},b^{\prime})$ must satisfy the inequality
\begin{align}
         \vert S\left(a,b,a', b'\right) \vert \le 2 
         \label{CHSH}
\end{align}
 for any choice of measurement settings $(a, a’, b, b’)$. Here, $M(\alpha,\beta)$ is the average product, $M(\alpha,\beta) =  
     \sum_{X_A,X_B} X_AX_BP(X_A,X_B\lvert\alpha,\beta) $,
where $P(X_A,X_B\vert \alpha,\beta)$ is the joint probability of measurements $(X_A, X_B)$ when the left and right analyzers are set to ($\alpha$, $\beta$). We note that Eq.~\ref{CHSH}) is cast in the form of the CHSH inequality~\cite{Clauser}.


It has been well established that bipartite quantum systems can violate inequality~\ref{CHSH} for a judicious choice of ($a$, $a ^{\prime}$, $b$, $b^{\prime})$,
with a maximum value $S$ = 2$\sqrt{2}$ corresponding to the Tsirelson bound.
Bell tests were first performed with static analyzer settings~\cite{Aspect1982a}, so could not strictly rule out signalling between the two subsystems. Specifically, the left measurement $X_A$ could in principle be influenced by the right analyzer setting $\beta$ (or, similarly, $X_B$ by $\alpha$)
through long-range interactions between the two subsystems. Such `static' Bell tests have also been performed with pairs of massive entangled particles ~\cite{Rowe_Bellions}.
Technological advances in quantum optics and electronics have enabled this `locality' loophole to be closed via dynamic Bell tests~\cite{Aspect1982b,Weihs,Scheidl2010,Hensen2015, Brunner_Review, Gisin_PRL, Giustina}, in which the detector
settings $\alpha$ and $\beta$ are altered just prior to measurement, so that the two measurement events are
space-like separated. Considerable effort has been made to closing other loopholes, including the ‘detection’ loophole (that posits that the detection efficiency depends on ($\alpha$, $\beta$)) and the
measurement-independence loophole, through a series of Bell tests of increasing sophistication and precision~\cite{Ansmann2009,Hofmann2012,Giustina2013,Hensen2015,Rauch}.  Nevertheless, Morgan~\cite{Morgan2006} and Vervoort~\cite{Vervoort2018} have questioned whether the measurement-independence loophole can be closed in systems with a background medium, wherein the hidden field variables may be influenced by the analyzer settings.
With a view to addressing this question, we here present a new platform for conducting Bell tests in the pilot-wave hydrodynamic system. 
Specifically, we devise and execute a static Bell test on the walking-droplet system, as a first step towards a dynamic test.

\begin{figure}
\includegraphics[width=0.5\textwidth]{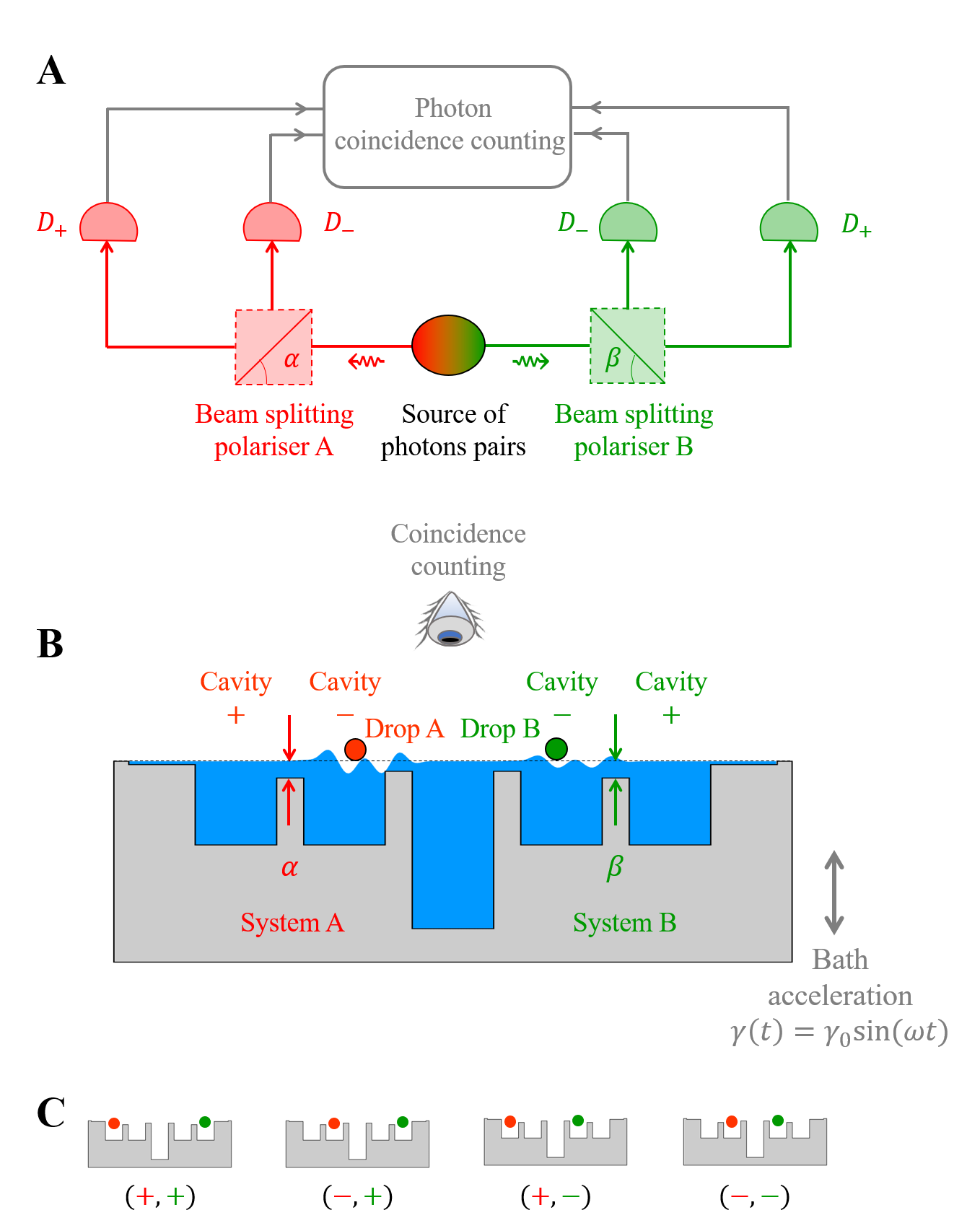}
\caption{{\bf Photonic and hydrodynamic Bell test arrangements}. {\bf A)} Schematic of an optical Bell test setup. Pairs of entangled photons are produced at the source and sent in opposite directions to two measurement stations, conventionally named A and B. At each station, one of two measurement (polarizer) settings is selected randomly and the measurement is performed. The measurement outcomes, which can take on two possible values, +1 or -1, are noted.
{\bf B)} Schematic of our hydrodynamic Bell test. The system consists of a pair of drops (red and green) walking on the surface of a vibrating liquid bath (blue) that spans the solid substrate (grey). Each drop is confined to its subsystem, a pair of wells separated by barriers across which they may tunnel unpredictably at a rate influenced by the barrier depths $\alpha$ and $\beta$, as may assume values of $a, a'$ or $b,b'$, respectively.
{\bf C)} The four possibilities for the drop configuration space, $(X_A,X_B)$, in the hydrodynamic Bell test.} 
\label{fig:Fig1}
\end{figure}

We consider a pair of walking droplets in the bipartite tunneling system introduced by Papatryfonos {\it et al.}~\cite{papatryfonos} in their demonstration of a hydrodynamic analog of quantum superradiance (See Figure 1a). The two subsystems, labelled A and B, correspond to single, wave-generating particles confined to a pair of identical cavities separated by a barrier across which the particles may tunnel. Each particle generates waves and moves in response to them according to the mathematical model of walking droplets detailed in the Methods section. In each subsystem, the preferred cavity corresponds to the ground state $|->$ and the other to the excited state $|+>$. The ground state
may be either the inner or outer cavity, depending on the length of the outer cavities~\cite{papatryfonos}. For the specific geometry considered here, the inner cavity of each subsystem is the ground state $|->$; the outer cavity, the excited state, $|+>$. 

The two subsystems are separated by a coupling cavity of variable length $L_c$, and by barriers that are sufficiently high as to preclude the particles from tunneling into the coupling cavity. Waves are transmitted across the central cavity, and so provide the coupling between subsystems A and B. The efficiency of this coupling is prescribed by the geometry of the central cavity: by increasing its depth $d_c$, the coupling may be increased, allowing the coupling cavity to serve as a nearly resonant transmission line ~\cite{Nachbin2018}. Transitions between ground and excited states in the subsystems correspond to individual tunneling events, the rate of which 
depends on the depths of the left and right barriers, and the 
length of the coupling cavity, $L_c$. As illustrated in Fig. 1a and 1c, we identify the state ($|+>$ or $|->$) of each subsystem with the dichotomic property $X$ in the optical Bell test, and the depths of the left and right barriers with the analyzer settings (respectively, $\alpha$ and $\beta$). The four possible combinations of $(X_A,X_B)$ are shown in Fig. 1c. We proceed by employing the numerical simulation method developed by Nachbin~\cite{Nachbin2017,Nachbin2018} to identify conditions under which Bell’s inequality is violated. 

To collect statistics in a manner comparable to the optical Bell tests with entangled photons, we proceed as follows. Each run begins by placing the two particles at random positions within their own subsystem. Their trajectories are then calculated for 2000 Faraday periods. At that time, measurements are made in each subsystem.
Specifically, we note the cavity in which the particle is located, and assign $X$ the value of $+1$ if the particle is in the outer cavity or $-1$ if it is in the inner cavity. The methodology for collecting the data is detailed in the Methods section. 
As in our recent study of hydrodynamic superradiance 
~\cite{papatryfonos}, the barrier depths $(\alpha, \beta)$ prescribe the local tunneling probability, and may also influence that of the distant partner drop. We proceed by demonstrating that for judicious choice of pairs of measurement settings $(\alpha, \beta)$, Bell's inequality may be violated. Our main result is shown in Fig. 2, indicating a narrow parameter range in which the CHSH inequality is violated. 
In the remainder of this section we will analyze in detail how this violation comes about. 

While the Bell inequality can be violated in our system with four different values of the measurement settings ($a$, $a’$, $b$, $b’$), our exploration of the ($a$, $a’$, $b$, $b’$) parameter-space indicates that a local maximum of $S$ arises when $b = a$ and $b’ = a’$, the symmetric case in which one may write $S(a, b, a’, b’) = S(a, a, a’, a’) = M(a, a) –  M(a’, a’) + 2M(a, a’)$. We deduced a maximum violation of $S_{\mathrm{max}}$ = 2.49 $\pm 0.04$ when $a$ = $b$ = $a^*$ = 0.099 cm and $a’$ = $b’$ = $a^{\prime*}$ = 0.1033 cm. In Fig. 2a, we plot $S$ as a function of $a’$ for fixed $a = a^*$, with the dashed line showing the limit $S$ = 2 above which the CHSH inequality (Eq.~\ref{CHSH}) is violated. While the inequality is violated only for a narrow range of parameters settings, in this parameter regime, the violation is clear, and the statistical confidence of the violation is above 20 standard deviations.  This behavior is reminiscent of the quantum case, where, without guidance from the theory, it is relatively difficult to find analyzer settings that allow for violation of the CHSH inequality, but for judiciously chosen settings, the inequality is violated substantially. 
Figures 2b and 2c show a typical example of the convergence of the `running average' with the number of runs which determines the relative error of our statistics.
This approach indicates when our statistics have converged for each $M(\alpha, \beta)$ calculation, specifically when the relative error has fallen below the prescribed tolerance.\\
\\
We proceed by detailing the manner in which Bell's inequality is violated in this classical pilot-wave system.
The maximum $S$ value occurs for moderate barrier depths, for which the droplets may become strongly correlated through the background wave field. In Fig. 3a, we show typical trajectories for the three combinations of measurement settings $(\alpha, \beta) \in \left\{(a^*,a^*),(a^*,a^{\prime*}),(a^{\prime*},a^{\prime*})\right\}$ that maximize $S$. For $(a,a’)=(a^{*},a^{\prime*})$, $S$ is maximized because $M(a^*, a^*)$ and $M(a^*, a^{\prime*})$ are large (see Fig. 3a top and middle panels), while  $M(a^{\prime*},a^{\prime*})$ is relatively small (Fig. 3a lower panel).
Fig. 3b corresponds to a shallow barrier, $a’$ = 0.0937 cm (the left-most value in Fig. 2a) and Fig. 3c to a relatively deep barrier, $a’$ = 0.11 cm (the right-most value in Fig. 2a). Figures 3b and c correspond to minima of $S$ occurring when the $a’$ barrier is either too shallow (Fig. 3b) or too deep (Fig. 3c).

The degree of synchronization in the droplet tunneling depends on the extent to which the droplets are affected by the barrier depth in the distant station. When the barrier depth in one station is too small,  the local particle is prevented from tunneling, regardless of the barrier depth in the other. The synchronization of states is thus reduced substantially. Conversely, when the barrier depth is too large, the particle generally tunnels across it, unaffected by the distant particle. Thus, the synchronization again remains relatively low. For intermediate barrier depths, each particle tunnels with a moderate probability that is strongly affected by the behavior of its distant partner. (We note that when only one drop is present, the variation of the measurement setting in the other subsystem does not influence its behavior. See Supplementary Information, Figure S1.)  As in our previous study of superradiance~\cite{papatryfonos}, the wave-mediated correlation creates a collective behavior of the droplet pairs. In particular, when one of the droplets tunnels to its excited state, the probability of the second droplet doing likewise increases substantially. Thus, through its wave-mediated interaction with its partner, each droplet is affected by the barrier depth of the distant station. 
\begin{figure}
\includegraphics[width=0.9\columnwidth]{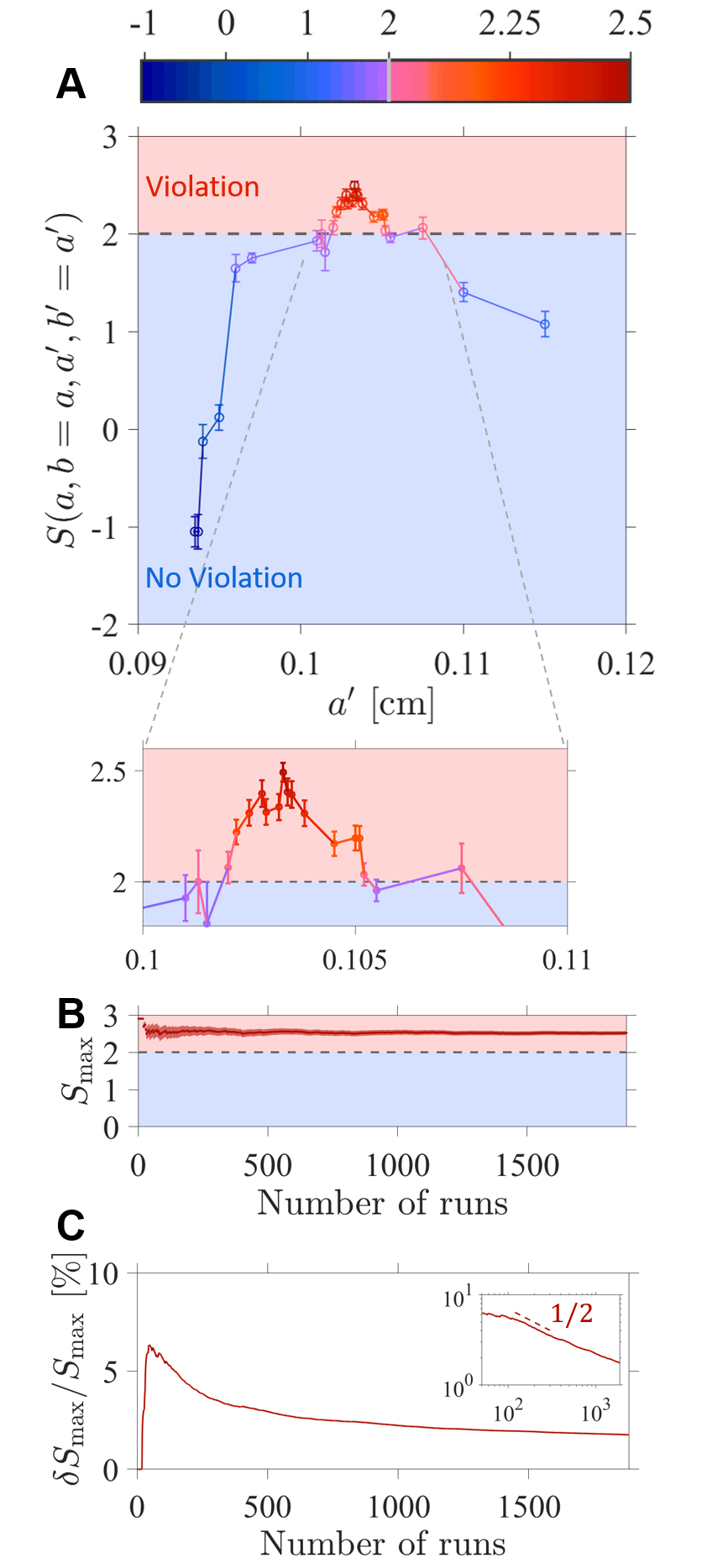}
\caption{Violation of Bell's inequality. {\bf A)} Bell parameter $S(a,a,a',a')$ as a function of the barrier depth, $a'$, for the symmetric case of $a=b$, $a' = b'$. For the calculation of the corresponding correlation functions, $M(a,a)$, $M(a, a')$ and $M(a',a')$, the barrier depth $a = a^*= 0.099$ cm remains fixed. For each combination of measurement settings, runs continue until statistics converge. The maximum Bell violation appears at $a^{\prime *} = 0.1033$cm, where $S = 2.49\pm 0.04$.  {\bf B)} Typical  curve showing the convergence of $S_{\max}$, the Bell parameter taken at the maximum point of violation $(a=b=a^*=0.099\;\mathrm{cm} ; a'=b'=a^{\prime*}=0.1033\;\mathrm{cm})$. 
The error bars indicate $\pm 3$ standard deviations. {\bf C)} Relative error $\delta S_{\max}/S_{\max}$ of the estimation of the Bell parameter $S$ with the number of runs, evaluated for the maximum point of violation $(a=b=a^* ; a'=b'=a^{\prime *})$. Inset: log-log scale; the dashed line indicates a $-1/2$ slope as expected from the convergence of an ensemble average.}
\label{fig:Fig2}
\end{figure}

\begin{figure}
\includegraphics[width=\columnwidth]{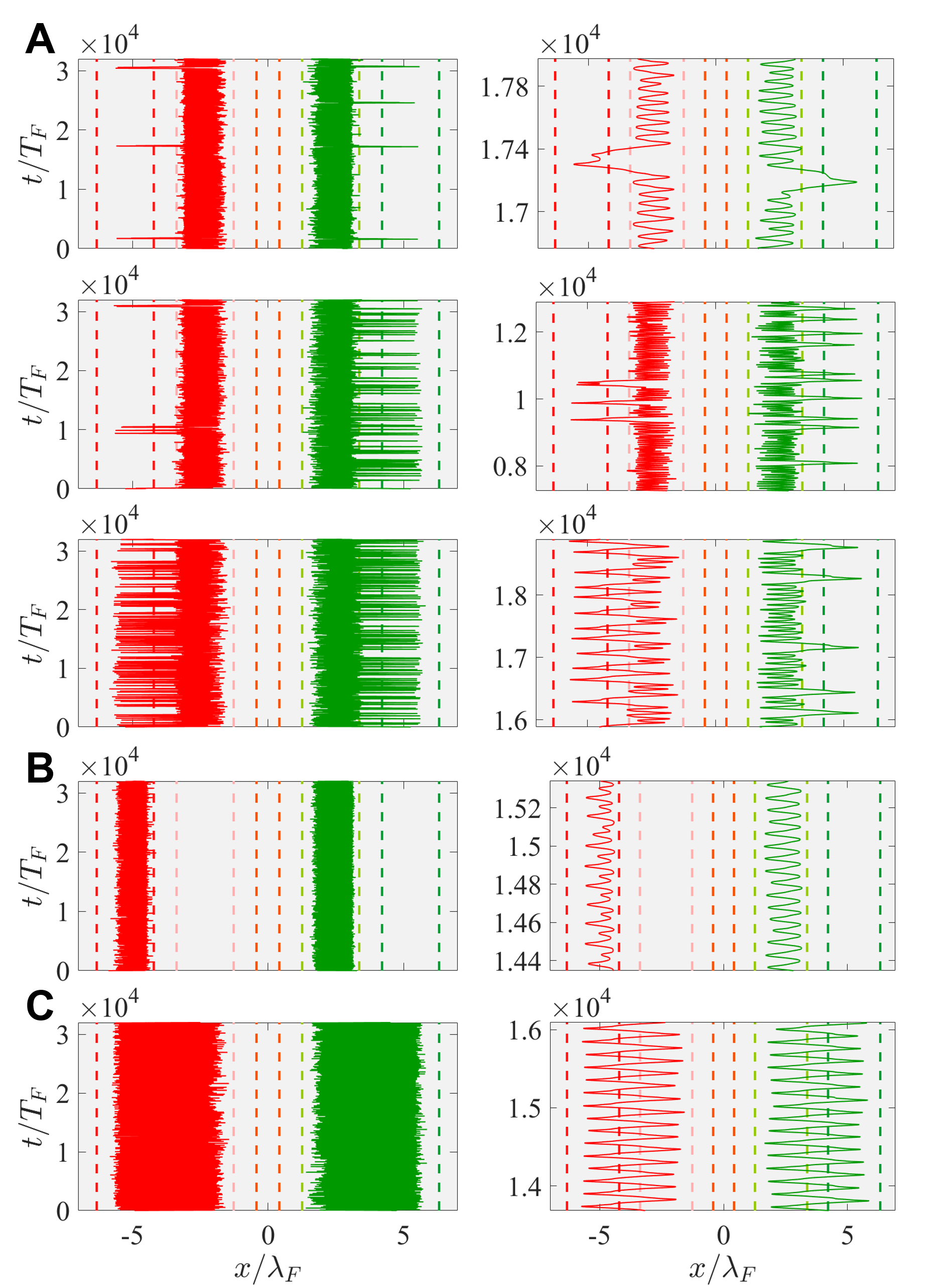}
\caption{{\bf Trajectory analysis}.  {\bf A-C)} Droplet trajectories for the symmetric case $(\alpha$,$\beta)$ $\in (a, a’)$  with $a=a^*$. Time evolves in the vertical direction. In {\bf (A)}, $a’=a^{\prime *}$=0.1033 cm (the maximizing value for $S$); in {\bf (B)}, $a’$ = 0.0937 cm; in {\bf (c)}, $a’$ = 0.11 cm. {\bf A)} Trajectories corresponding to the three correlation functions $M(a^*, a^*)=0.94$ (upper panel), $M(a^*, a^{\prime *})=0.84 $ (middle panel), and $M(a^{\prime *}, a^{\prime *})=0.13$ (lower panel). The tunneling events are highly correlated only in the upper and lower panels. {\bf B)} Trajectories corresponding to $M(a^*, a')$ with $a'= 0.0937$ cm. When the barrier depth $a’$ is sufficiently small, the wave-mediated communication between droplets is diminished, and droplets tend to get trapped in one cavity, leading to minima of $S$ and $M(a^*, a’)\approx 0$ when we average over the droplet's initial conditions. {\bf C)} Another minimum of $M(a^*, a’)$ and $S$ occurs when one of the barrier depths is too large, in which case one of the droplets tunnels continuously, unimpeded by the barrier, as if it were in a single cavity. Averaging over all initial conditions leads to a relatively low value of $M(a^*, a’)\approx 0$. Note that the correlation function is deduced by averaging over all initial conditions in both subsystems.}
\label{fig:Fig3}
\end{figure}


Non-factorizable states arise in multi- and bipartite systems when the state of the whole cannot be simply defined in terms of the state of its subsystems~\cite{Classical_entanglement,Horodecki09}. A canonical example is the singlet state of entangled photons. In our system, non-separability manifests itself through the fact that the joint probability of the two dichotomic states $X_A$ and $X_B$, specifically $P(X_A,X_B\vert \alpha,\beta)$, is not equal to the product $P(X_A \vert \alpha) P(X_B \vert \beta)$. 
Classical non-factorizable states have been demonstrated in both electromagnetic and acoustic wave systems~\cite{spreeuw_classical_1998,Classical_entanglement}, and also in single-particle systems, for example through consideration of the internal degrees of freedom of a single atom~\cite{Classical_entanglement}. However, neither the classical wave states nor the internal degrees of freedom in the single-particle system can be spatially separated~\cite{spreeuw_classical_1998,Classical_entanglement}. Thus,
such classical states can be used for neither performing dynamic Bell tests nor building analogs to qubits and quantum computing ~\cite{spreeuw_classical_1998,Classical_entanglement}. 
Conversely, our bipartite system exhibits spatially separated non-factorizable states, and so introduces the possibility of exploring novel forms of quantum-inspired classical computing, and performing more sophisticated Bell tests.

We have devised a platform for performing static Bell tests on a classical bipartite pilot-wave system. The maximum violation was found to be $2.49\pm 0.04$, and arose when the system geometries were chosen such that the droplet motion was marked by strongly synchronized tunneling for one measurement setting combination, moderate and weak synchronization for the others. A key step in the process was recognizing that the system geometry may serve as a proxy for analyzer settings. 
In our system, the barrier depths play the role of polarizer angles in the photonic Bell tests. We expect this conceptual advance to prompt and facilitate the numerical and experimental execution of Bell tests in pilot-wave hydrodynamics and other stochastic bipartite classical systems.



Long-range correlations have been reported in other bipartite hydrodynamic pilot-wave systems, most notably in the establishment of statistical indistinguishability of a pair of distant droplets~\cite{Nachbin2018} and in the recent analog of superradiance~\cite{papatryfonos}. In these examples, if observers were unaware of the pilot wave field and observed only the droplets, they could only account for the observed correlations by inferring a nonlocal connection between the droplet pairs. 
The violation of Bell’s Inequality (Eq.~\ref{CHSH}) in our static Bell test may be rationalized in terms of the wave-mediated coupling between the two subsystems. Specifically, the form of the pilot wave and the concomitant particle positions depend on the geometry of both subsystems, and so necessarily violate the assumption of Bell locality.
The feasibility of adapting our platform to incorporate time-dependent topography has recently been demonstrated by Nachbin~\cite{Nachbin2022}.
This adaptation will allow for future Bell tests in which communication between the two subsystems may be eliminated and analyzer settings changed dynamically.

\section*{Methods}
\subsection*{Numerical method}
System parameters are chosen to correspond to a fluid bath of density 0.95 g/cm$^3$, viscosity 16 cS and surface tension 20.9 dynes/cm vibrating vertically in a sinusoidal fashion with peak acceleration $\gamma_0$
and frequency $\omega_0 = 80$Hz. The resonant bouncing of the particle at the Faraday frequency triggers a quasi-monochromatic damped wave pattern with a corresponding 
Faraday wavelength of $\lambda_F=4.75$~mm.
Each of the four cavities has a fixed length of 1.2 cm, corresponding to approximately 2.5$\lambda_F$.
In all simulations, we set the coupling cavity depth to $d_c=6.3~\lambda_F $ which ensures strong inter-cavity coupling. 
We thus describe our bipartite tunneling system in terms of two coupled, two-level systems, as shown schematically in Figure~\ref{fig:Fig1}b. 

\begin{figure*} 
\includegraphics[width=\textwidth]{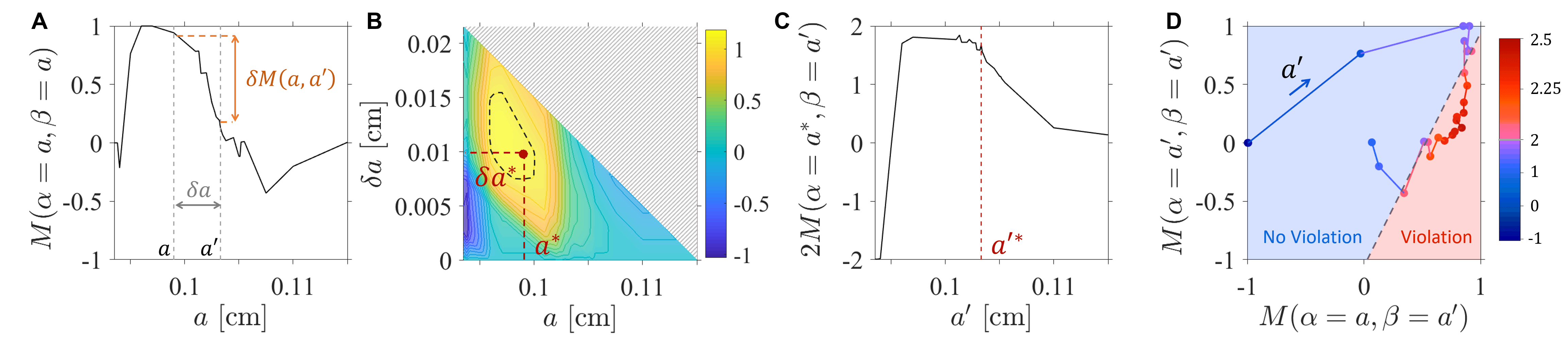}
\caption{{\bf Strategy to optimize \bm{$S(a,a,a',a')=M(\alpha=a,\beta=a)-M(\alpha= a^\prime,\beta=a^\prime)+2M(\alpha=a,\beta=a')$} }. $S$ is optimized by searching a parameter regime $(a,a')$ near the maximum of $\delta M(a,a')=M(\alpha = a,\beta= a)-M(\alpha=a',\beta=a')$ and a range of $a'$ that maximizes $M(a,a')$ with $a$ fixed. {\bf A)} Evolution of $M(\alpha=a, \beta =a)$ as a function of $a$. The indicated values $a^*$ = 0.099 cm and $a^{\prime *}$ = 0.1033 cm are the S-maximizing values used in Figs. 2-3. Note $\delta a= a'-a$. The difference between the corresponding correlation functions $\delta M(a,a+\delta a)=\delta M(a,a')=M(\alpha=a,\beta=a)-M(\alpha=a',\beta=a')$ is marked in orange. {\bf B)} Optimization of $\delta M(\alpha=a, \beta =a+\delta a)$ as a function of barrier depth $a$ and $\delta a $.  The domain for which $(\max_{a,a'} (\delta M)-\delta M)/\max_{a,a'}(\delta M)>0.9$ is bound by the black dashed lines. {\bf C)} 2$M(\alpha=a^*,\beta =a')$ as a function of depth $a’$ for fixed $a = a^*$. {\bf D)} Evolution of $S (a,a,a',a')$ in the correlation representation space $(M(a,a');M(a',a'))$ with $a=a^*$. The direction of increasing $a'$ is indicated by the blue arrow. The dots are colored with respect to their $S$ values. A grey dashed line indicates the limiting case $S=2$. }   
\label{fig:Fig4}
\end{figure*}

Nachbin et al. \cite{Nachbin2017,Nachbin2018} formulated a theoretical model for the one-dimensional motion of walking droplets over a vibrating liquid bath with complex topography. Here we adjust this model in order to consider the cooperative tunneling of two identical particles in the geometry depicted in Figure \ref{fig:Fig1}b. The positions, $x_j$ ($j=1,2$), of the two identical particles of mass $m$ evolve according to Newton's Law:
\be
m\ddot{x}_j+ c~F(t)\dot{x}_j = - F(t)\; \frac{\partial\eta}{\partial x}(x_j(t),t).
\label{Drop1ODE}
\ee
The particle moves in response to gradients of the wave elevation $\eta(x,t)$, which thus plays the role of a time-dependent potential. The particle motion is resisted by a drag force proportional to its speed. The drag constant $c$ follows from the modeling presented in Molacek \& Bush \cite{Molacek2013b}.
The time dependence of these propulsive and drag forces is prescribed by $F(t)$, as arises in the walker system owing to the droplet's bouncing~\cite{Bush2015a,Milewski2015,Nachbin2017}. In terms of their lateral motion, the particles are viewed as horizontal oscillators that can transition unpredictably between two neighboring cavities. The dichotomic property $X$ assessed for Bell’s inequality is assigned according to the particle location $x_j = x_j(t_m)\; (j = A,B)$ at the measurement time $t_m$.
Specifically, $X = +1$ if the drop is in the outer, excited state, and $X = -1$ if it is in the inner, ground state.

The particles serve as moving wave sources that establish their own time-dependent wave potential that is computed as follows. The velocity potential of the liquid bath $\phi(x,z,t)$ is a harmonic function satisfying Laplace's equation. In the bulk of the fluid, the velocity field is given by $(u,v)=\nabla\phi$.  
The wave model is formulated in the bath's reference frame, where the effective gravity is $g(t)= g+\gamma_0 \sin(\omega_0 t)$, where $g$ is the gravitation acceleration, and {$\gamma_0$} is the amplitude of the bath's vibrational acceleration. The wave field thus evolves according to \cite{Milewski2015,Nachbin2017}: 
\be
\frac{\partial \eta}{\partial t} = \frac{\partial \phi}{\partial z} + 2\nu 
\frac{\partial^2\eta}{\partial x^2},
\label{Kin}
\ee
\be
\frac{\partial \phi}{\partial t}  =  - g(t) \eta + \frac{\sigma}{\rho} \frac{\partial^2\eta}{\partial x^2} + 2\nu 
\frac{\partial^2\phi}{\partial x^2} -\!\! \sum_{j=1,2} \frac{P_d(x-x_j(t))}{\rho}.
\label{Bern}
\ee
The particles ($j=A,B$) generate waves on the free surface by applying local pressure terms $P_d$. The wave forcing term $P_d(x-x_j(t))$ and the coefficient $F(t)$ are activated 
only during a fraction of the Faraday period $T_F$, corresponding to the contact time $T_c$ in the walking-droplet system and approximated by $T_c=T_F/4$. The particle is assumed to be in resonance with the
most unstable (subharmonic) Faraday mode of the bath~\cite{Milewski2015}, a key feature of pilot-wave hydrodynamics \cite{Protiere2006,Bush2015a,BushOza}. The numerical approach to simulating Eqs. (2)-(3) is detailed in the Supplementary Information.

\subsection*{Measurement procedure and data collection}
To initialize the runs, the wave and velocity fields of the bath are set to zero, and the particle positions are assigned random, uniformly distributed values. Then, the model runs for 2000 Faraday periods, a measurement is made, and all fields are reset back to zero to initialize the subsequent run. 
This cycle is repeated for each set of parameter settings until the relative error in the running average of $M(\alpha = a,\beta = b)$ is reduced to an acceptably small value. We set this tolerance to be 3$\%$ for parameters that violate the inequality and 7$\%$ for those that do not. While extremely accurate, this `discrete' technique is computationally intensive; thus
we have used it only for the most critical points of the parameter space, in which the maximal Bell violations occurred. 
To explore the parameter space more efficiently, we adopt an alternative, relatively expedient, `continuous' approach, in which the final conditions of one run serve as the initial conditions of the next.
We demonstrated the statistical equivalence of the two approaches as follows. For specific selected data points, we performed approximately 30 different runs using the two techniques, and found the results of the ‘discrete’ and ‘continuous’ runs to be in agreement to within 3$\%$. 
We then executed continuous runs for 48,000 Faraday periods, during which measurements are performed frequently at uniformly distributed random times. After a sufficiently long run, the full range of initial conditions will have been effectively explored. The consistency of the results deduced with the discrete and continuous approaches demonstrates that 
the long-time emergent statistics are independent of the initial conditions. 

Since the inequality involves four
different correlation functions (three for the symmetric case considered here), finding the combinations of measurement settings that maximized $S$ was not entirely straightforward. Figure 4 summarises the strategy we followed in seeking violations. We first investigated the evolution of a single correlation function $M(\alpha =a, \beta =a)$ as a function of $a$. This gave us a good sense of parameters that maximize the difference $\delta M(a,a')=M(a,a)-M(a',a')$ (see Figure 4a). $\delta M(a,a')$ involves two of the correlation functions of Eq. 1, in the symmetric case of interest where $a$=$b$ and $a’$=$b’$. Figure 4b shows a 2D plot of the optimisation of $\delta M$ as a function of $a$ and $a'$. The black dashed lines highlight the domain in which $(\max_{a,a'} (\delta M)-\delta M)/\max_{a,a'}(\delta M)>0.9$. The other term in the inequality, specifically $2M(\alpha=a,\beta=a')$, represents a combination of measurements from unequal barrier depths at the two measurement stations. Figure 4c represents the dependence of 2$M(\alpha =a^*, \beta =a’)$ on depth $a’$ for fixed $a=a^*$, the $S$-maximizing value considered in Figures 2 and 3. Finally, Figure 4d shows the evolution of the correlation functions $(M(a,a');M(a',a'))$ with increasing $a'$ and fixed $a=a^*$. \\

\vspace*{0.1in}

\section{Data availability}
The data that support the findings of our study are available upon request.

{\section{Code Availability}}

{The code that generated the data is available upon request.}

\bibliographystyle{unsrt}
\bibliography{ARFMBib}

\begin{thebibliography}{10}

\bibitem{Couder2005a}
Y.~Couder, S.~Proti\`{e}re, E.~Fort, and A.~Boudaoud.
\newblock Walking and orbiting droplets.
\newblock {\em Nature}, 437(208), 2005.

\bibitem{Couder2006}
Y.~Couder and E.~Fort.
\newblock Single particle diffraction and interference at a macroscopic scale.
\newblock {\em Phys. Rev. Lett.}, 97(154101), 2006.

\bibitem{Fort2010}
E.~Fort, A.~Eddi, J.~Moukhtar, A.~Boudaoud, and Y.~Couder.
\newblock Path-memory induced quantization of classical orbits.
\newblock {\em Proc. Natl. Acad. Sci.}, 107(41):17515--17520, 2010.

\bibitem{Bush2015a}
J.~W.~M. Bush.
\newblock Pilot-wave hydrodynamics.
\newblock {\em Ann. Rev. Fluid Mech.}, 47, 2015.

\bibitem{BushOza}
J.~W.~M. Bush and A.~U. Oza.
\newblock Hydrodynamic quantum analogs.
\newblock {\em Reports on Progress in Physics}, 84(1):017001, 2020.

\bibitem{Pucci2018}
G.~Pucci, D.~M. Harris, L.~M. Faria, and J.~W.~M. Bush.
\newblock Walking droplets interacting with single and double slits.
\newblock {\em J. Fluid Mech.}, 835:1136--1156, 2018.

\bibitem{Ellegaard2020}
C.~Ellegaard and M.~T. Levinsen.
\newblock Interaction of wave-driven particles with slit structures.
\newblock {\em arXiv:2005.12335}, 2020.

\bibitem{Perrard2014}
S.~Perrard, M.~Labousse, M.~Miskin, E.~Fort, and Y.~Couder.
\newblock Self-organization into quantized eigenstates of a classical
  wave-driven particle.
\newblock {\em Nat. Commun.}, 5(3219), 2014.

\bibitem{Eddi2009b}
A.~Eddi, E.~Fort, F.~Moisy, and Y.~Couder.
\newblock Unpredictable tunneling of a classical wave-particle association.
\newblock {\em Phys. Rev. Lett.}, 102(240401), 2009.

\bibitem{Saenz2019a}
P.~J. S\'{a}enz, T.~Cristea-Platon, and J.~W.~M. Bush.
\newblock A hydrodynamic analog of {F}riedel oscillations.
\newblock {\em Science Advances}, 6(20), 2020.

\bibitem{saenz_emergent_2021}
P.~J. Sáenz, G.~Pucci, S.~E. Turton, A.~Goujon, R.~R. Rosales, J.~Dunkel, and
  J.~W.~M. Bush.
\newblock Emergent order in hydrodynamic spin lattices.
\newblock {\em Nature}, 596(7870):58--62, August 2021.

\bibitem{Harris2013a}
D.~M. Harris, J.~Moukhtar, E.~Fort, Y.~Couder, and J.~W.~M. Bush.
\newblock Wavelike statistics from pilot-wave dynamics in a circular corral.
\newblock {\em Phys. Rev. E}, 88(011001):1--5, 2013.

\bibitem{Saenz2018b}
P.~J. S\'{a}enz, G.~Pucci, A.~Goujon, T.~Cristea-Platon, J.~Dunkel, and
  J.~W.~M. Bush.
\newblock Spin lattices of walking droplets.
\newblock {\em Phys. Rev. Fluids}, 3(100508), 2018.

\bibitem{Eddi2011a}
A.~Eddi, E.~Sultan, J.~Moukhtar, E.~Fort, M.~Rossi, and Y.~Couder.
\newblock Information stored in {F}araday waves: the origin of a path memory.
\newblock {\em J. Fluid Mech.}, 674:433--463, 2011.

\bibitem{Harris2018}
D.~M. Harris, P.-T. Brun, A.~Damiano, L.~Faria, and J.~W.~M. Bush.
\newblock The interaction of a walking droplet and a submerged pillar: {F}rom
  scattering to the logarithmic spiral.
\newblock {\em Chaos}, 28(096106), 2018.

\bibitem{Nachbin2018}
A.~Nachbin.
\newblock Walking droplets correlated at a distance.
\newblock {\em Chaos}, 28(096110), 2018.

\bibitem{Nachbin2022}
A.~Nachbin.
\newblock The effect of isolation on two-particle correlations in pilot-wave
  hydrodynamics.
\newblock {\em Phys. Rev. Fluids}, to appear, 2022.

\bibitem{papatryfonos}
K.~Papatryfonos, M.~Ruelle, C.~Bourdiol, A.~Nachbin, J.~W.~M. Bush, and
  M.~Labousse.
\newblock Hydrodynamic superradiance in wave-mediated cooperative tunneling.
\newblock {\em Communications Physics}, 5:142, 2022.

\bibitem{DeVoe}
R.~G. DeVoe and R.G. Brewer.
\newblock Observation of superradiant and subradiant spontaneous emission of
  two trapped ions.
\newblock {\em Phys. Rev. Lett.}, 76:2049--2052, Mar 1996.

\bibitem{makarov_metastable_2004}
A.~A. Makarov and V.~S. Letokhov.
\newblock Metastable entangled states of atomic systems in macroscale:
  radiation dynamics and spectrum.
\newblock {\em International Workshop on Quantum Optics 2003 (Proceedings of
  the SPIE)}, pages 54--64, 2004.
\newblock Publisher: SPIE.

\bibitem{Kaminer}
A.~Karnieli, N.~Rivera, A.~Arie, and I.~Kaminer.
\newblock Superradiance and subradiance due to quantum interference of
  entangled free electrons.
\newblock {\em Phys. Rev. Lett.}, 127:060403, Aug 2021.

\bibitem{TANJISUZUKI2011201}
H.~Tanji-Suzuki, I.~D. Leroux, M.~H. Schleier-Smith, M.~Cetina, A.~T. Grier,
  J.~Simon, and V.~Vuletić.
\newblock Chapter 4 - interaction between atomic ensembles and optical
  resonators: Classical description.
\newblock In E.~Arimondo, P.R. Berman, and C.C. Lin, editors, {\em Advances in
  Atomic, Molecular, and Optical Physics}, volume~60 of {\em Advances In
  Atomic, Molecular, and Optical Physics}, pages 201--237. Academic Press,
  2011.

\bibitem{de2015emerging}
L.~De~la Pe{\~n}a, A.~M. Cetto, and A.~Vald{\'e}s-Hern{\'a}ndez.
\newblock The emerging quantum.
\newblock {\em The Physics behind Quantum Mechanics. Cham: Springer
  International Publishing}, 2015.

\bibitem{Vervoort2018}
L.~Vervoort.
\newblock Are hidden-variable theories for pilot-wave systems possible?
\newblock {\em Foundations of Physics}, 48:803--826, 2018.

\bibitem{Bell1964}
J.~S. Bell.
\newblock On the {E}instein {P}odolsky {R}osen paradox.
\newblock {\em Physics}, 1(3):195--200, 1964.

\bibitem{Einstein1935}
A.~Einstein, B.~Podolsky, and N.~Rosen.
\newblock Can quantum-mechanical description of physical reality be considered
  complete?
\newblock {\em Physical Review}, 47(777), 1935.

\bibitem{Clauser}
J.F. Clauser, M.A. Horne, A.~Shimony, and R.A. Holt.
\newblock Proposed experiment to test local hidden-variable theories.
\newblock {\em Phys. Rev. Lett.}, 23:880--4, 1969.

\bibitem{Aspect1982a}
A.~Aspect, P.~Grangier, and G.~Roger.
\newblock Experimental realization of {E}instein-{P}odolsky-{R}osen-{B}ohm {\it
  gedankenexperiment}: A new violation of {B}ell's inequalities.
\newblock {\em Phys. Rev. Lett.}, 49(91), 1982.

\bibitem{Rowe_Bellions}
M.~A. Rowe, D.~Kielpinski, V.~Meyer, C.~A. Sackett, W.~M. Itano, C.~Monroe, and
  D.~J. Wineland.
\newblock Experimental violation of a {Bell}'s inequality with efficient
  detection.
\newblock {\em Nature}, 409(6822):791--794, February 2001.
\newblock Number: 6822 Publisher: Nature Publishing Group.

\bibitem{Aspect1982b}
A.~Aspect, J.~Dalibard, and G.~Roger.
\newblock Experimental test of {B}ell's inequalities using time-varying
  analyzers.
\newblock {\em Phys. Rev. Lett.}, 49(1804), 1982.

\bibitem{Weihs}
G.~Weihs, T.~Jennewein, C.~Simon, H.~Weinfurter, and A.~Zeilinger.
\newblock Violation of bell's inequality under strict einstein locality
  conditions.
\newblock {\em Phys. Rev. Lett.}, 81:5039--5043, Dec 1998.

\bibitem{Scheidl2010}
T.~Scheidl, R.~Ursin, J.~Kofler, S.~Ramelow, X.~Ma, T.~Herbst, L.~Ratschbacher,
  A.~Fedrizzi, N.~K. Langford, T.~Jennewein, , and A.~Zeilinger.
\newblock Violation of local realism with freedom of choice.
\newblock {\em PNAS}, 107:19708--19713, 2010.

\bibitem{Hensen2015}
B.~Hensen, H.~Bernien, A.~Dr\'eau, A.~Reiserer, N.~Kalb, M.~S. Blok,
  J.~Ruitenberg, R.~F.~L. Vermeulen, R.~N. Schouten, C.~Abell\'an, W.~Amaya,
  V.~Pruneri, M.~W. Mitchell, M.~Markham, D.~J. Twitchen, D.~Elkouss,
  S.~Wehner, T.~H. Taminiau, and R.~Hanson.
\newblock Loophole-free bell inequality violation using electron spins
  separated by 1.3 kilometres.
\newblock {\em Nature}, 526:682--686, 2015.

\bibitem{Brunner_Review}
N.~Brunner, D.~Cavalcanti, S.~Pironio, V.~Scarani, and S.~Wehner.
\newblock Bell nonlocality.
\newblock {\em Rev. Mod. Phys.}, 86:419--478, Apr 2014.

\bibitem{Gisin_PRL}
W.~Tittel, J.~Brendel, H.~Zbinden, and N.~Gisin.
\newblock Violation of bell inequalities by photons more than 10 km apart.
\newblock {\em Phys. Rev. Lett.}, 81:3563--3566, Oct 1998.

\bibitem{Giustina}
M.~Giustina, M.~A.~M. Versteegh, S.~Wengerowsky, J.~Handsteiner, A.~Hochrainer,
  K.~Phelan, F.~Steinlechner, J.~Kofler, J.-A. Larsson, C.~Abell\'an, W.~Amaya,
  V.~Pruneri, M.~W. Mitchell, J.~Beyer, T.~Gerrits, A.~E. Lita, L.~K. Shalm,
  S.~W. Nam, T.~Scheidl, R.~Ursin, B.~Wittmann, and A.~Zeilinger.
\newblock Significant-loophole-free test of bell's theorem with entangled
  photons.
\newblock {\em Phys. Rev. Lett.}, 115:250401, Dec 2015.

\bibitem{Ansmann2009}
M.~Ansmann, H.~Wang, R.~C. Bialczak, M.~Hofheinz, E.~Lucero, M.~Neeley, A.~D.
  O'Connell, D.~Sank, M.~Weides, J.~Wenner, Cleland~A. N., and Martinis~J. M.
\newblock Violation of bell's inequality in josephson phase qubits.
\newblock {\em Nature}, 461:504--506, 2009.

\bibitem{Hofmann2012}
J.~Hofmann, M~Krug, N.~Ortegel, L.~G\'erard, M.~Weber, Rosenfeld W., and
  Weinfurter H.
\newblock Heralded entanglement between widely separated atoms.
\newblock {\em Science}, 337:72--75, 2012.

\bibitem{Giustina2013}
M.~Giustina, A.~Mech, S.~Ramelow, B.~Wittmann, J.~Kofler, J.~Beyer, A.~Lita,
  B.~Calkins, T.~Gerrits, S.~W. Nam, R.~Ursin, and A.~Zeilinger.
\newblock Bell violation using entangled photons without the fair-sampling
  assumption.
\newblock {\em Nature}, 497:227--230, 2013.

\bibitem{Rauch}
D.~Rauch, J.~Handsteiner, A.~Hochrainer, J.~Gallicchio, A.~S. Friedman,
  C.~Leung, B.~Liu, L.~Bulla, S.~Ecker, F.~Steinlechner, R.~Ursin, B.~Hu,
  D.~Leon, C.~Benn, A.~Ghedina, M.~Cecconi, A.~H. Guth, D.~I. Kaiser,
  T.~Scheidl, and A.~Zeilinger.
\newblock Cosmic bell test using random measurement settings from high-redshift
  quasars.
\newblock {\em Phys. Rev. Lett.}, 121:080403, Aug 2018.

\bibitem{Morgan2006}
P.~Morgan.
\newblock Bell inequalities for random fields.
\newblock {\em J. Phys. A}, 39:7441--7455, 2006.

\bibitem{Nachbin2017}
A.~Nachbin, P.~A. Milewski, and J.~W.~M. Bush.
\newblock Tunneling with a hydrodynamic pilot-wave model.
\newblock {\em Physical Review Fluids}, 2(034801), 2017.

\bibitem{Classical_entanglement}
E.~Karimi and R.~W. Boyd.
\newblock Classical entanglement?
\newblock {\em Science}, 350(6265):1172--1173, December 2015.

\bibitem{Horodecki09}
R.~Horodecki, P.~Horodecki, M.~Horodecki, and K.~Horodecki.
\newblock Quantum entanglement.
\newblock {\em Rev. Mod. Phys.}, 81:865--942, 2009.

\bibitem{spreeuw_classical_1998}
R.~J.~C. Spreeuw.
\newblock A {Classical} {Analogy} of {Entanglement}.
\newblock {\em Foundations of Physics}, 28(3):361--374, March 1998.

\bibitem{Molacek2013b}
J.~Mol\'{a}\v{c}ek and J.~W.~M. Bush.
\newblock Drops walking on a vibrating bath: towards a hydrodynamic pilot-wave
  theory.
\newblock {\em J. Fluid Mech.}, 727:612--647, 2013.

\bibitem{Milewski2015}
P.~Milewski, C.~Galeano-Rios, A.~Nachbin, and J.~W.~M. Bush.
\newblock Faraday pilot-wave dynamics: modelling and computation.
\newblock {\em J. Fluid Mech.}, 778:361--388, 2015.

\bibitem{Protiere2006}
S.~Proti{\`e}re, A.~Boudaoud, and Y.~Couder.
\newblock Particle-wave association on a fluid interface.
\newblock {\em J. Fluid. Mech.}, 554:85--108, 2006.

\end{thebibliography}
\end{document}